\documentclass[twocolumn,pra,preprintnumbers,amsmath,amssymb]{revtex4}
\usepackage{graphicx}
\usepackage{latexsym}
\usepackage{color}
\begin{document}

\title{Static traversable wormholes in Lyra manifold}
\author{A. Sayahian Jahromi$^1$, H. Moradpour$^2$\footnote{h.moradpour@riaam.ac.ir}}
\address{$^1$ Zarghan Branch,Islamic Azad University, Zarghan, Iran\\
$^2$ Research Institute for Astronomy and Astrophysics of Maragha
(RIAAM), P.O. Box 55134-441, Maragha, Iran}

\begin{abstract}
\noindent At first, considering the Einstein framework, we
introduce some new static traversable wormholes, and study the
effects of a dark energy-like source on them. Thereinafter, a
brief review on Einstein field equations in Lyra manifold is
presented, and we address some static traversable wormholes in the
Lyra manifold which satisfy the energy-conditions. It is also
shown that solutions introduced in the Einstein framework may also
meet the energy conditions in the Lyra manifold. Finally, we focus
on vacuum Lyra manifold and find some traversable asymptotically
flat wormholes. In summary, our study shows that it is
theoretically possible to find a Lyra displacement vector field in
a manner in which traversable wormholes satisfy the energy
conditions in a Lyra manifold.
\end{abstract}

\maketitle
\section{Introduction}

Gravitational theories open a window to the interstellar trips
through the support of notion of traversable wormhole
\cite{Wheeler1,Wheeler2,Thorne1,Thorne2,Visser1,cut-paste1,prds,r,Lobo2,Lobo1,Oliveira,Sajadi,Garcia2,Garcia3,loboaip,nw,cr1,cr2}.
Besides, the physics of wormholes is so close to that of black
holes
\cite{Hayward,Kardashev,Kuhfittig,Sushkov-Zaslavskii,Cai1,Cai2}, a
point which also encourages us to study wormholes. Although, in
the Einstein theory, traversable wormholes do not respect all
energy conditions simultaneously \cite{cut-paste1}, there are some
gravitational theories in which a traversable wormhole may support
the energy conditions
\cite{prds,r,Lobo2,Lobo1,Oliveira,mehdi,kord,mehdi1,mehdic,tr1,tr2,tr3,tr4,tr5,tr6,msad}.
In fact, find various traversable wormholes in different
gravitational theories is an important and attractive issue in
theoretical physics
\cite{msad,Sushkov,Lobo2005,Lobo20051,lpr,pr,hrm,hrmc,chack,cut-paste2,cut-paste3,cut-paste4,cut-paste5,cut-paste6,cut-paste7,cut-paste8,cut-paste9,nw1}.

In the Einstein theory of gravity, the spacetime is a Riemannian
manifold, geometry and energy-momentum sources are coupled to each
other in a minimal way, and the spacetime is curved by the
energy-momentum source \cite{poisson}. In one hand, the successes
of Einstein theory confirm that the spacetime is not flat
everywhere \cite{poisson}. On the other hand, due to the failure
of this theory to provide proper description for various
phenomenons, physicists try to modify it
\cite{beyond,beyond1,cmc,cmc1,cmc2,rastall,mhds}.

Using the Lyra generalization of the Riemannian geometry
\cite{L1}, Sen has constructed a modified version for the Einstein
theory \cite{sen}, which attracts a lot of attention to itself
\cite{dv1,dv2,lrev,seng,D1,s1,s2,hara,lyraobs,lkur,lyrather,ln1,sepangi,ln2}.
Some conformal dynamic traversable wormholes have also been
studied in this theory which violate the null and weak energy
conditions \cite{dwhl}.

Traversable wormholes should satisfy the flaring-out condition
\cite{Thorne2,cut-paste1}. In the Einstein framework, this
condition leads to the fact that the sum of the energy density and
the radial pressure is not positive everywhere, and thus the
energy conditions are not met by traversable wormholes in the
Einstein framework\cite{Thorne2,cut-paste1}. Here, we are
interested to investigate the possibility of satisfying the energy
conditions by static spherically symmetric traversable wormholes
in the Lyra manifold.

In order to achieve our goal, at first, the general mathematical
properties of static traversable wormholes are addressed in the
next section. In addition, we introduce some new static
traversable wormholes in the Einstein framework, and study the
effects of a dark energy-like source on the solutions in
Sec.~($\textmd{III}$). In fact, the provisions of this section
also help us in understanding some difficulties of traversable
wormholes in the Einstein framework. In the fourth section,
providing a brief review on the Einstein field equations in the
ordinary gauge of the Lyra manifold, we show that some solutions
obtained in Sec.~($\textmd{III}$) can meet the energy conditions.
More solutions are also studied in this section. In
Sec.~($\textmd{V}$), considering the empty Lyra manifold, the
possibility of obtaining traversable wormholes as solutions of
vacuum field equations is also investigated.
Section~($\textmd{VI}$) is devoted to a summary.

\section{Traversable wormholes, general remarks\label{ter}}

The general spherically symmetric traversable wormhole metric is
written as

\begin{equation}\label{met}
ds^{2}=-U(r)dt^{2}+\frac{dr^{2}}{1-\frac{b\left(r\right)}{r}}+r^{2}[d\theta^{2}+\sin^{2}\theta
d\phi^{2}],
\end{equation}

\noindent where $b(r)$ and $U(r)$ are the shape and redshift
functions, respectively. In order to have a wormhole with a throat
located at $r_0$, $b(r)$ should satisfy the $b(r_0)=r_0$
condition. Besides, the metric has to preserve its signature for
$r> r_0$ meaning that we must have $U(r)>0$ and $b(r)<r$ when
$r>r_0$. In addition, it is useful to note here that to avoid
getting horizon instead of the wormhole's throat, the redshift
function should satisfy the $U(r_0)>0$ condition. Finally, the
satisfaction of the flaring-out condition ($f(r_0)\equiv
b^{\prime}(r_0)-1<0$) is necessary for obtaining a traversable
wormhole \cite{Thorne2}.

Now, bearing the Einstein field equations in mind
($G_{\mu\nu}=T_{\mu\nu}$), one reaches

\begin{eqnarray}\label{den}
\rho&=&\frac{b^{\prime}(r)}{r^2},\\ \nonumber
p_r&=&\frac{U^{\prime}
(r)}{r U(r)}(1-\frac{b(r)}{r})-\frac{b(r)}{r^3},\\
\nonumber p_t&=&p_r+\frac{r}{2}\left[p_r^{\prime}+\left(\rho
+p_r\right)\frac{U^{\prime}(r)}{2U(r)}\right],
\end{eqnarray}

\noindent for the non-zero components of the energy-momentum
source that supports the geometry which is a spherically symmetric
spacetime. In the above equation and also in this paper, prime
denotes the derivative with respect to $r$. Considering the
flaring-out as well as $b(r_0)=r_0$ conditions, one can use
Eq.~(\ref{den}) to obtain $\rho(r_0)+p_r(r_0)<0$ in general
relativity (GR) \cite{Thorne2}. In fact, the components of
energy-momentum tensor should satisfy the

\begin{eqnarray}\label{ec}
WEC&\rightarrow&\ \rho\geq0,\ \rho+p_i>0,\nonumber\\
NEC&\rightarrow&\ \rho+p_i\geq0,\nonumber\\
DEC&\rightarrow&\ \rho\geq0,\ \rho\geq|p_i|,\nonumber\\
SEC&\rightarrow&\ \rho+p_i\geq0,\ \rho+p_r+2p_t\geq0,
\end{eqnarray}

\noindent conditions \cite{poisson}, where WEC and NEC are the
weak energy condition and the null energy condition, respectively.
DEC also denotes the dominant energy condition which declares that
the velocity of energy transfer cannot be more than that of light.
SEC, used as the abbreviation of the strong energy condition,
stems from the attractive nature of gravity, and its form is the
direct result of considering a spherically symmetric metric in the
GR framework \cite{poisson}. Therefore, it is apparent that, since
$\rho(r_0)+p_r(r_0)<0$, traversable wormholes do not respect these
conditions everywhere \cite{cut-paste1}.

Here, we should remind that the above form of SEC is the direct
result of working in the GR framework, and if one uses another
gravitational theory, then SEC differs from the above relation
\cite{poisson}. It is also useful to remind that the energy
conditions may be obtained by traversable wormholes in some
modified gravity
\cite{prds,r,Lobo2,Lobo1,Oliveira,mehdi,kord,mehdi1,mehdic,tr1,tr2,tr3,tr4,tr5,tr6,msad}.

Since anisotropy is arisen in various astrophysical phenomena,
such as different types of wormholes and stars, etc.
\cite{as1,as2,aw1,aw6,aw10,as3,as4,as5,aw4,aw3,aw7,as6,as7,aw5,aw9,as8,as9,as10,as12,as13,as11,as14},
we are going to investigate this quantity in our study. One can
also find some features of isotropic static spherically symmetric
wormholes in Ref.~\cite{iw}. The anisotropy parameter is defined
as

\begin{eqnarray}\label{an1}
\Delta=p_t-p_r,
\end{eqnarray}

\noindent which finally leads to

\begin{eqnarray}\label{an2}
\Delta=\frac{r}{2}\left[p_r^{\prime}+\left(\rho
+p_r\right)\frac{U^{\prime}(r)}{2U(r)}\right],
\end{eqnarray}

\noindent for the above metric. Geometry is attractive (repulsive)
for $p_t<p_r$ ($p_t>p_r$). The $\Delta=0$ case also means that an
isotropic fluid supports the geometry.

\section{Traversable wormholes with hyperbolic shape function}

\begin{figure}[ht]
\centering
\includegraphics[width=2in, height=2in]{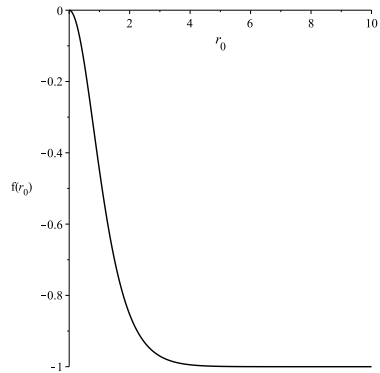}
\caption{\label{f1} The plot depicts $f(r_0)$ function.}
\end{figure}

Let us consider $b(r)=a\tanh(r)$ where $a$ is a constant. The
$b(r_0)=r_0$ and flaring-out conditions imply
$a=\frac{r_0}{\tanh(r_0)}$ and
$f(r_0)=\frac{r_0}{\tanh(r_0)}(1-\tanh^2(r_0))-1<0$, respectively.
In this situation, we have $b(r)<r$ for $r>r_0$ and
$b(r)/r\rightarrow0$ at the $r\rightarrow\infty$ limit. As it is
obvious from Fig.~(\ref{f1}), the flaring-out condition is
respected by wormholes with arbitrary values of $r_0$.

\subsection{Wormholes with constant redshift function}\label{11}

\begin{figure}[ht] \centering
\includegraphics[width=2in, height=2in]{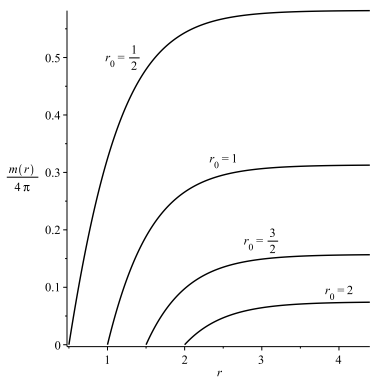}
\caption{\label{m1} The function $\frac{m(r)}{4\pi}$ for some
values of $r_0$.}
\end{figure}

As the first example, we set $U(r)=V$. Therefore, without lose of
generality, one can set $U(r)=1$ meaning that the metric is
asymptotically flat. Following Eq.~(\ref{den}), one reaches at

\begin{eqnarray}\label{en01}
\rho&=&\frac{r_0[1-\tanh^2(r)]}{r^2\tanh(r_0)},\nonumber\\
p_r&=&-\frac{b(r)}{r^3},\\
\nonumber
w_r(r)&=&\frac{\tanh(r)}{r[\tanh^2(r)-1]},\\
\nonumber
p_t&=&\frac{r_0[r(\tanh^2(r)-1)+\tanh(r)]}{2r^3\tanh(r_0)}
\end{eqnarray}

\noindent for the energy density and pressure components of the
fluid which supports this geometry. As we have already seen, due
to the flaring-out condition, the sum of energy density and radial
pressure is of key importance in studying traversable wormholes.
Therefore, considering this point as well as the spherical
symmetry of metric, we calculated the radial state parameter
$w_r(r)(\equiv\frac{p_r}{\rho})$ in the above equation which helps
us in simplifying our analysis. Now, defining the mass function as

\begin{eqnarray}\label{massfunc}
m(r)\equiv\int_{r_0}^r \rho 4\pi r^2dr,
\end{eqnarray}

\noindent we easily obtain

\begin{eqnarray}\label{mass1}
m(r)=\frac{4\pi r_0}{\tanh(r_0)}[\tanh(r)-\tanh(r_0)],
\end{eqnarray}

\noindent for mass which is independent of the redshift function.
$\frac{m(r)}{4\pi}$ as the function of $r$ has been plotted in
Fig.~(\ref{m1}) for different values of $r_0$. As it is apparent,
at the $r\rightarrow\infty$ limit, we have $m\approx\frac{4\pi
r_0}{\tanh(r_0)}[1-\tanh(r_0)]$ which is clearly a positive
confined quantity. Therefore, an observer, located at infinity,
measures a confined and positive mass for wormhole satisfying
Eq.~(\ref{mass1}).

For this solution, $\Delta\geq0$ for $r\geq0$ signalling us a
repulsive geometry. In addition, we have $\rho>0$, $p_t\geq0$ and
thus $\rho+p_t>0$ for all values of $r$. Besides, although $\rho$
is positive, $w_r(r)<-1$ for $r>0$ meaning that the
$\rho+p_r\geq0$ condition is violated everywhere. Therefore, there
is no radial interval in which this solution can satisfy the
energy conditions.

\subsection{Solutions with exponential redshift function}\label{nn}

Here, in order to obtain more traversable wormholes, we consider a
more general case for the redshift function as $U(r)=v\exp(cr^n)$,
where $v$, $c$ and $n$ are unknown constants evaluated later. It
is apparent that the $c=0$ case leads to the redshift function
$U(r)=constant$ studied in previous section~(\ref{11}). In this
manner, calculations for the Einstein field equations lead to

\begin{eqnarray}\label{en1}
\rho&=&\frac{r_0[1-\tanh^2(r)]}{r^2\tanh(r_0)},\\
p_r&=&cnr^{n-2}-\frac{b(r)}{r^3}[cnr^n+1],\nonumber\\
\nonumber
w_r(r)&=&\frac{[cnr^{n+1}-(\frac{r_0\tanh(r)}{\tanh(r_0)})(cnr^n+1)]\tanh(r_0)}{rr_0[1-\tanh^2(r)]},\\
\nonumber
p_t&=&\frac{1}{4r^3\tanh(r_0)}[(cnr^n)^2(r\tanh(r_0)-r_0\tanh(r))\nonumber\\
&+&cnr^n(r_0(r\tanh^2(r)+(1-2n)\tanh(r))\nonumber\\
&+&r(2n\tanh(r_0)-r_0))\nonumber\\
&+&2r_0(r(\tanh^2(r)-1)+\tanh(r))].\nonumber
\end{eqnarray}

\begin{figure}[ht] \centering
\includegraphics[width=2in, height=2in]{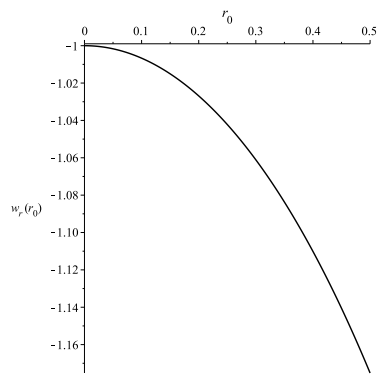}
\caption{\label{omega1} The plot depicts $w_r(r_0)$.}
\end{figure}

\noindent It is also easy to check that the result obtained in
Sec.~(\ref{11}) is recovered at the appropriate limit of
$c\rightarrow0$. Moreover, as it is obvious from~(\ref{den}),
Eq.~(\ref{mass1}) is also valid in this case meaning that they are
solutions with bounded mass function. Using Eq.~(\ref{en1}), one
gets

\begin{eqnarray}\label{omega}
w_r(r_0)=\frac{\tanh(r_0)}{r_0[\tanh^2(r_0)-1]},
\end{eqnarray}

\noindent which is a relation between the wormhole throat and the
value of the radial state parameter at the wormhole throat. One
can also get this result by using Eq.~(\ref{en01}) which belongs
to the $U(r)=1$ case. It is the direct result of this fact that,
at throat, we have $b(r_0)=r_0$ and thus
$p(r_0)+\rho(r_0)=\frac{b^\prime(r_0)-1}{r_0^2}$ which is
independent of the redshift function~(\ref{den}). $w_r$ has been
plotted as a function of $r_0$ in Fig.~(\ref{omega1}) indicating
that $w_r(r_0)<-1$ for $r_0>0$, in full agreement with the
flaring-out condition.


\subsection*{Asymptotically flat wormholes\label{int}}

\begin{figure}[ht]
\centering
\includegraphics[width=2.5in, height=2.5in]{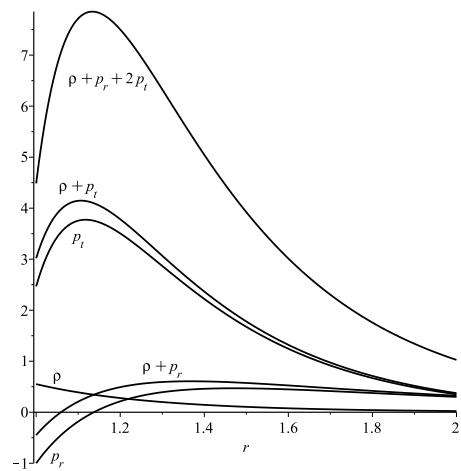}
\caption{\label{fe1} All the curves tend to zero at the
$r\rightarrow\infty$ limit which is in accordance with the
asymptotically flat condition. It is obvious that energy
conditions are not met by this solution.}
\end{figure}

In order to have the asymptotically flat wormholes, we should have
$U(r)\rightarrow1$ and $b(r)/r\rightarrow0$ at the
$r\rightarrow\infty$ limit. We have previously mentioned that the
$b(r)/r\rightarrow0$ condition is met by $b(r)=a\tanh(r)$.
Besides, applying the asymptotically flat condition to the
redshift function, one finds that, independent of the value of
$c$, if $n<0$ and $v=1$, then the asymptotically flat condition is
preserved by the redshift function. Moreover, since $p_r(r_0)<0$,
in order to have positive radial pressure at radius $r>R$, the
radial pressure should satisfy the $p_r(R)=0$ condition leading to

\begin{eqnarray}\label{c1}
c=\frac{b(R)R^{-n-1}}{n(1-\frac{b(R)}{R})},
\end{eqnarray}

\noindent for $c$. Based on this result, since $n<0$, the positive
radial pressure is accessible only for negative $c$, and thus, we
only consider this case. In addition, this equation shows that, at
the $R\rightarrow r_0$ limit, if $b(R)=r_0+\epsilon$, then we have
$c\approx\frac{1+\epsilon}{n\epsilon
r_0^n}[1-(n+1)\frac{\epsilon}{r_0}]$. Therefore, choosing very
small values for $\epsilon$, one can miniaturize the radial
interval wherein the radial pressure is negative, but it is
impossible to eliminate this interval by putting $\epsilon=0$
leading to $c\rightarrow-\infty$ which is meaningless. This result
is fully compatible with the flaring-out condition. In
Fig.~(\ref{fe1}), density and pressure components as well as the
weak, null and strong energy conditions have been plotted as the
functions of $r$ for $n=-2$, $c=-10$ and $r_0=1$. Choosing smaller
values for $r_0$ and $c$, one can minimize the interval in which
the $\rho+p_r\geq0$ condition is violated. It is also useful to
note that due to the flaring-out condition ($b^\prime(r_0)-1<0$)
as well as the $b(r_0)=r_0$ condition, leading to $p_r(r_0)<0$,
this interval cannot completely be disappeared. In addition, the
$\rho+p_r\geq0$ condition starts to be satisfied for
$r\geq\tilde{R}$, where $\tilde{R}$ is the radius for which
$\rho+p_r=0$. This yields the

\begin{eqnarray}\label{c2}
c=\frac{b(\tilde{R})(\tilde{R})^{-n-1}-b^\prime(\tilde{R})\tilde{R}^{-n}}{n(1-\frac{b(\tilde{R})}{\tilde{R}})},
\end{eqnarray}

\noindent relation. Comparing this result with Eq.~(\ref{c1}), one
can obtain $R\neq\tilde{R}$ consistent with Fig.~(\ref{fe1}). It
is also interesting to mention here that if we considered the
$b(r)=r_0$ case, leading to $\rho=b^\prime(r)=0$ (or equally
$b^\prime(\tilde{R})=0$), then we faced $R=\tilde{R}$ meaning that
both the $\rho+p_r\geq0$ and $p_r\geq0$ conditions start to be met
from the same radius.

Finally, it is useful to note that, unlike radial pressure, the
energy density of wormholes studied in Secs.~(\ref{11})
and~(\ref{nn}) is positive everywhere. This result is in full
agreement with the confined positive mass obtained in these
sections meaning that the mass content of the total
energy-momentum source supporting such geometries is positive and
confined. We also found out that the energy conditions are not met
by the obtained wormholes.

\subsection{Isotropic wormholes}

\begin{figure}
  \includegraphics[width=2.2in, height=2.2in]{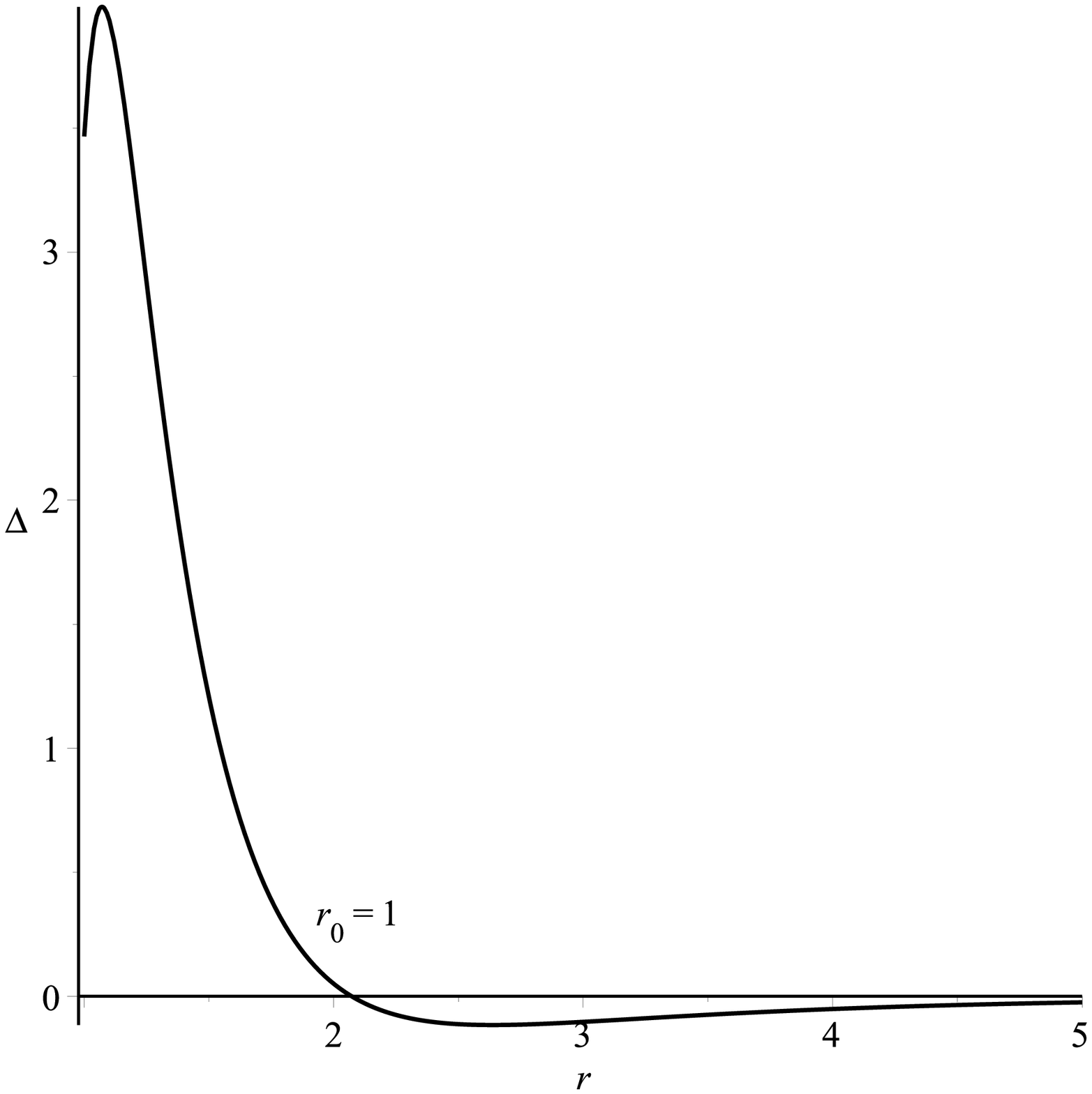}\ \ \
  \includegraphics[width=2.2in, height=2.2in]{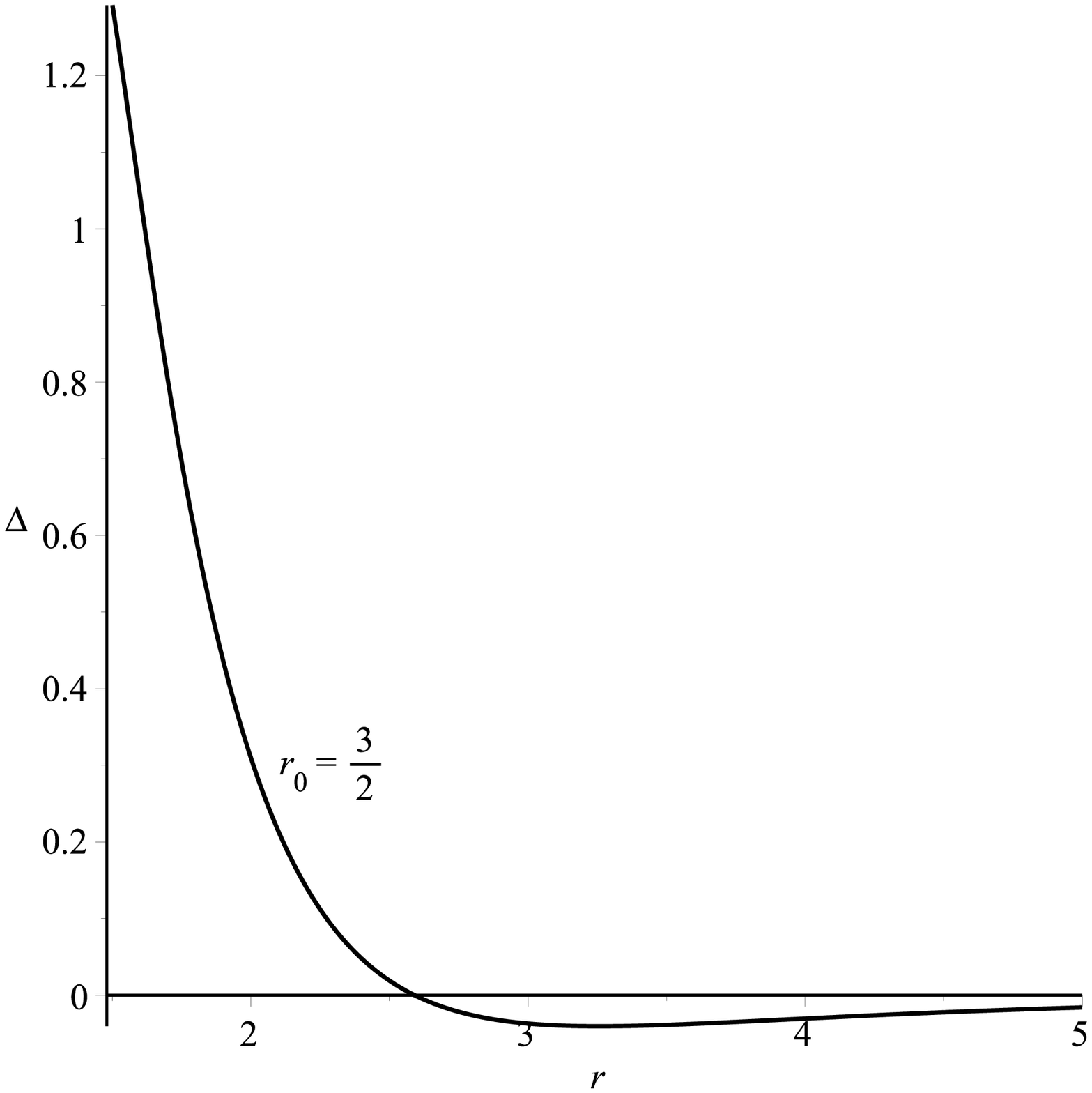}
 \caption{\label{an1t}
The $\Delta$ parameter as a function of $r$ for some values of
$r_0$, whenever $n=-2$, $c=-10$. In accordance with the
asymptotically flat behavior of metric, we have
$\Delta\rightarrow0$ for $r\rightarrow\infty$.}
\end{figure}

Before addressing some isotropic traversable wormholes, we study
the behavior of anisotropy parameter of the case studied in
Sec.~(\ref{nn}). The anisotropy parameter is plotted as a function
of radius for some values of $r_0$, when $n=-2$ and $c=-10$, in
Fig.~(\ref{an1t}). As it is obvious from both Eq.~(\ref{en1}) and
Fig.~(\ref{an1t}), the anisotropy is vanished at the
$r\rightarrow\infty$ limit. Moreover, it is also apparent that
there is a transition from a repulsive geometry to an attractive
geometry in both studied cases.

From Eq.~(\ref{an1}), in order to obtain the isotropic solutions
($p_t=p_r$), we should have $\Delta=0$ that yields

\begin{eqnarray}\label{an3}
\frac{U^{\prime}(r)}{2U(r)}=-\frac{p_r^{\prime}}{(\rho + p_r)}.
\end{eqnarray}

\noindent One may use this equation to find the redshift function,
if the radial state parameter is identified. Generally, one needs
to know the $p_r(\rho)$ relation for solving this equation. Here,
we only consider the $w_r=constant$ case. For a fluid of constant
$w_r$, this equation leads to

\begin{eqnarray}\label{an4}
U(r)=U_0\rho^\frac{-2w_r}{1+w_r},
\end{eqnarray}

\noindent and thus

\begin{eqnarray}\label{an5}
U(r)=U_0(\frac{r_0[1-\tanh^2(r)]}{r^2\tanh(r_0)})^\frac{-2w_r}{1+w_r},
\end{eqnarray}

\noindent where we have used Eq.~(\ref{en1}) to obtain this
equation. It is easy to check that Eq.~(\ref{omega}) is also valid
in this case meaning that we have
$w_r=\frac{\tanh(r_0)}{r_0[\tanh^2(r_0)-1]}<-1$ for $r_0>0$.
Therefore, independent of the value of $r_0$, we have
$U(r\rightarrow\infty)\rightarrow\infty$ meaning that these
isotropic traversable wormholes are not asymptotically flat.

Bearing Eq.~(\ref{an3}) in mind, it is obvious that solutions with
$U(r)=constant$ may be obtained if we have either $w_r=0$ or
$p_r=\frac{\rho}{w_r}=p_0\equiv constant$. Based on the above
argument, and in agreement with the flaring-out condition, the
$w_r=0$ case does not lead to traversable wormholes and thus it is
forbidden. For the second case, using Eq.~(\ref{den}), one can
easily get $b(r)=-p_0r^3$ which does not clearly respect the
$b(r)<r$ condition for $r>r_0$. In addition, since the throat
condition ($b(r_0)=r_0$) implies $p_0=-\frac{1}{r_0^2}$, it is
apparent that this case does not respect the flaring-out
condition. Therefore, this case does not lead to wormhole, a
result in agreement with our study in Sec.~(\ref{11}). Further
details on isotropic wormholes in the GR framework can be found in
Ref.~\cite{iw}.


\subsection{Hyperbolic wormholes in the presence of a dark energy-like source}

Dark energy, as a controversial problem in physics, may be modeled
by an ideal prefect fluid with $\rho=-p=constant>0$. Since the
current accelerating universe is dominated by dark energy, we are
going to study the effects of this cosmic fluid on solutions found
out in Sec.~(\ref{nn}). Moreover, we also point out to the
$\rho<0$ case used to describe an anti de-Sitter universe. In
order to achieve this goal, we consider a universe with
$U(r)=v\exp(cr^n)$ filled by two fluids. In our model, one of them
($\Lambda$ source) has constant energy density and pressure which
satisfy the $\rho_\Lambda=-p_\Lambda=\Lambda$ relation, and the
other has energy-momentum tensor $T_{\ \mu}^{x\
\nu}=diag(-\rho^x,p_r^x,p_t^x,p_t^x)$. It is also worth to remind
that the asymptotically flat condition implies $n<0$. In this
situation, the Einstein field equations are written as

\begin{eqnarray}\label{eec1}
G_{\alpha\beta}=T_{\alpha\beta}=T^x_{\alpha\beta}-\Lambda
g_{\alpha\beta}.
\end{eqnarray}
%
%
Using Eq.~(\ref{en1}), one reaches at

\begin{eqnarray}\label{en2}
\rho^x&=&\rho-\Lambda=\frac{r_0[1-\tanh^2(r)]}{r^2\tanh(r_0)}-\Lambda,\\
\nonumber p_r^x&=&p_r+\Lambda=cnr^{n-2}-\frac{b(r)}{r^3}[cnr^n+1]+\Lambda,\\
\nonumber
p_t^x&=&p_t+\Lambda=\frac{1}{4r^3\tanh(r_0)}[(cnr^n)^2(r\tanh(r_0)\nonumber\\
&-&r_0\tanh(r))+cnr^n(r_0(r\tanh^2(r)\nonumber\\&+&(1-2n)\tanh(r))+r(2n\tanh(r_0)-r_0))\nonumber\\&+&2r_0(r(\tanh^2(r)-1)+\tanh(r))]+\Lambda.\nonumber
\end{eqnarray}

\noindent Therefore, at the wormhole throat, we have

\begin{eqnarray}
w_r^x(r_0)=\frac{p_r^x}{\rho^x}=\frac{\Lambda-\frac{1}{r_0^2}}{\frac{1-\tanh^2(r_0)}{r_0\tanh(r_0)}-\Lambda},
\end{eqnarray}

\noindent which implies that $\rho^x(r_0)<0$, $p_r^x(r_0)\leq0$
and $w_r^x(r_0)\geq0$ whenever
$\frac{1-\tanh^2(r_0)}{r_0\tanh(r_0)}<\Lambda\leq\frac{1}{r_0^2}$.
In addition, for this range of the $\Lambda$ parameter, we have
$\rho^x(r)<0$ meaning that WEC cannot be satisfied by the $T_{\
\mu}^{x\ \nu}$ source. We should note that although the energy
density of total energy-momentum tensor ($\rho$) is positive
everywhere, the energy density of unknown fluid ($\rho^x$) may be
negative depending on the value of $\Lambda$. Moreover, it was
found out that unlike the total energy-momentum tensor, for which
$w_r(r_0)\leq-1$, one gets $w_r^x(r_0)\geq0$ for the unknown
fluid. In fact these state parameters are in a mutual relation as

\begin{eqnarray}
w_r(r)=w_r^x(r)(1-\frac{\Lambda}{\rho})+\frac{\Lambda}{\rho},
\end{eqnarray}

\noindent indicating that, at the $\Lambda\rightarrow0$ limit, we
have $w_r(r)\rightarrow w_r^x(r)$ compatible with our previous
results obtained in Sec.~(\ref{nn}).

\begin{figure}[ht] \centering
\includegraphics[width=3in, height=3in]{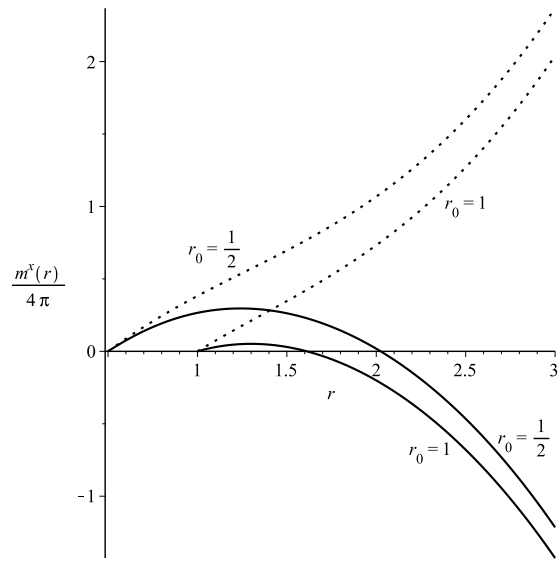}
\caption{\label{m2} The function $\frac{m^x(r)}{4\pi}$ for some
values of $r_0$. Dot and solid lines belong to the
$\Lambda=-\frac{1}{5}$ and $\Lambda=\frac{1}{5}$ cases,
respectively. Although $m^x(r)$ is divergent, the total mass
$m(r)$, satisfying Eq.~(\ref{mass1}), is bounded and positive.}
\end{figure}

It is also obvious that Eq.~(\ref{mass1}) holds for the total
energy-momentum tensor with energy density $\rho$ which is a
positive quantity (see Eq.~(\ref{en1})). Moreover, for the mass
corresponding to the unknown fluid ($m^x(r)$), we get

\begin{eqnarray}\label{mass2}
m^x(r)&=&\int_{r_0}^r \rho^x 4\pi r^2
dr=m(r)-\frac{4\pi\Lambda}{3}(r^3-r_0^3)\nonumber\\
&=&\frac{4\pi
r_0}{\tanh(r_0)}[\tanh(r)-\tanh(r_0)]\nonumber\\&-&\frac{4\pi\Lambda}{3}(r^3-r_0^3),
\end{eqnarray}

\noindent plotted for some values of $\Lambda$ and $r_0$ in
Fig.~(\ref{m2}). As it is obvious from both Fig.~(\ref{m2}) and
Eq.~(\ref{mass2}), for distances away from the wormhole throat
($r\gg r_0$), we have $m^x(r)\approx-\frac{4\pi\Lambda}{3}r^3$
which is negative (positive) for $\Lambda>0$ ($\Lambda<0$). The
latter is due to this fact that at the $r\rightarrow\infty$ limit,
we have $m\simeq C$ in accordance with the
$b(r\rightarrow\infty)\simeq C^{\prime}$ behavior, where $C$ and
$C^{\prime}$ are constant. Since the mass corresponding to the
$\Lambda$ source ($m^\Lambda(r)=\frac{4\pi\Lambda}{3}(r^3-r_0^3)$)
increases with increasing radius, at first glance, one could have
expected the total mass ($m(r)$) to be divergent. But,
Eq.~(\ref{mass2}) shows the unknown fluid cancels the effect of
$\Lambda$ on the total mass, indicating that the total mass
satisfies Eq.~(\ref{mass1}). In fact, this result can also be
obtained using Eq.~(\ref{en2}) which implies
$m^x(r)=m(r)-m^\Lambda(r)$. Therefore, the infinity observer gets
a positive and bounded mass for the wormholes embedded in a
spacetime filled by a source as $\rho_\Lambda=-p_\Lambda=\Lambda$
together with an unknown fluid which meets Eqs.~(\ref{en2})
and~(\ref{mass2}).

Eq.~(\ref{en2}) yields $\rho^x+p_r^x=\rho+p_r$, meaning that
$\Lambda$ does not affect the $\rho+p_r$ term, and hence, the
result obtained in Eq.~(\ref{c2}) is also valid here. In fact,
this is the direct result of the $\rho_\Lambda+p_\Lambda=0$
property of the $\Lambda$ source. In addition, due to the isotropy
of the $\Lambda$ source, one can expect
$\Delta=p_t-p_r=p_t^x-p_r^x$ confirmed by using Eq.~(\ref{en2}).
Additionally, the radial pressure of the unknown fluid gets its
positive values for $r\geq R_1$, where $R_1$ is evaluated from the
$p_r^x(R_1)=0$ condition leading to

\begin{eqnarray}\label{c3}
c=\frac{b(R_1)R_1^{-n-1}-\Lambda
R_1^{3-n}}{n(1-\frac{b(R_1)}{R_1})}.
\end{eqnarray}

\noindent Comparing this result with Eq.~(\ref{c2}), we find out
that both the $\rho^x+p_r^x=\rho+p_r\geq0$ and $p_r^x\geq0$
conditions may simultaneously be obtained if we have
$R_1=\tilde{R}$. In addition, a comparison between this result and
Eq.~(\ref{c1}) helps us in obtaining the effect of $\Lambda$ on
the interval for which $p_r^x(r)<0$. Indeed, one can use
Eq.~(\ref{en2}) in order to see that, unlike the $\rho+p_r$
expression, $\Lambda$ directly affects the intervals for which
$p_r^x(r)<0$ and $\rho_r^x(r)<0$. From Eq.~(\ref{en2}), it is also
obvious that $\rho^x(r)$ is positive whenever $\Lambda$ is
negative. Moreover, if $\rho(r_0)<\Lambda$, then we have
$\rho^x(r)<0$ for $r\geq r_0$. Finally, for the
$0<\Lambda<\rho(r_0)$ case, the energy density of unknown fluid is
negative if $r>\mathcal{R}$ in which $\mathcal{R}$ is evaluated
from the $\rho^x(\mathcal{R})=0$ condition as

\begin{eqnarray}\label{0c4}
\frac{1-\tanh^2(\mathcal{R})}{\mathcal{R}^2}=\frac{\Lambda\tanh(r_0)}{r_0}.
\end{eqnarray}

Using the above results as well as Eqs.~(\ref{c2}),~(\ref{en2})
and~(\ref{c3}), one obtains

\begin{eqnarray}\label{c4}
\frac{1-\tanh^2(R_1)}{R_1^3}=\frac{\Lambda\tanh(r_0)}{r_0}.
\end{eqnarray}

\noindent This equation establishes a relation between the radius
of wormhole throat, $\Lambda$ and $R_1$, and therefore, in order
to have a real solution for $R_1$, $\Lambda$ should be positive.
Therefore, the wormhole throat affects the value of $R_1$ and thus
the interval in which both the radial pressure ($p_r^x$) and the
$\rho^x+p_r^x$ quantity are negative. One can also combine
Eqs.~(\ref{0c4}) and~(\ref{c4}) with each other to get a mutual
relation between $\mathcal{R}$ and $R_1$ whenever
$0<\Lambda<\rho(r_0)$. In this manner, if $\mathcal{R}=R_1$, which
is available only for $\mathcal{R}=R_1=1$, then we have
$\rho^x+p_r^x=\rho+p_r\geq0$, $p_r^x\geq0$ and $\rho^x(r)<0$ for
$r>1$. As we saw, the existence of $\Lambda$ does not solve the
difficulty of violating energy conditions by obtained solutions,
and it may at most affect the rate of behavior of solutions.
Hence, we do not plot energy conditions in the presence of
$\Lambda$ in order to avoid prolongation of the article.

In summary, as we have seen in Sec.~(\ref{nn}), the total
energy-momentum tensor of obtained solutions does not respect the
energy conditions. Our study in this section also shows that the
problem of violation of energy conditions by traversable wormholes
cannot be solved by adding the $\Lambda$ term. In fact, although
the total mass of obtained solutions is positive, the behavior of
assumed unknown source does not represent a baryonic source.


\section{Lyra manifold and Einstein field equations}

Introducing the gauge function $H(x^\mu)$ into the structureless
manifold, and defining the displacement vector between two
neighboring points as $Q(x^\nu+dx^\nu)-q(x^\nu)\equiv
H(x^\mu)dx^\nu$, Lyra proposed a generalization to the Riemannian
geometry in which metric is invariant under the gauge and
coordinate transformations \cite{L1,sen}. Full details on this
geometry and its relation with Weyl geometry can be found in
\cite{L1,lrev}.

The Einstein gravitational field equations in the ordinary gauge
of the Lyra manifold, a generalization of the Riemannian geometry
\cite{L1}, are written as \cite{sen,D1}

\begin{eqnarray}\label{EinL}
\mathcal{G}_{\mu\nu}\equiv
G_{\mu\nu}+\tilde{G}_{\mu\nu}=T_{\mu\nu},
\end{eqnarray}

\noindent where

\begin{eqnarray}\label{lf}
\tilde{G}_{\mu\nu}=\frac{3}{2}\phi_\mu\phi_\nu-\frac{3}{4}\phi_\alpha\phi^\alpha
g_{\mu\nu},
\end{eqnarray}

\noindent and we considered $8\pi G=1=c$. Here, $\phi_\mu$ denotes
the Lyra displacement vector field which is the direct result of
introducing the gauge function $H(x^\mu)$ into the structureless
manifold \cite{L1,sen,lrev}. It is also useful to mention that for
spacetimes with diagonal $G_{\mu\nu}$ supported by a fluid with
energy-momentum tensor $T_\mu^\nu=diag(-\rho,p_r,p_t,p_t)$,
$\tilde{G}_{\mu\nu}$ should be diagonal meaning that only one
component of the vector field $\phi_\mu$ may not be zero, and its
three other components must be zero. It has been shown that the
$\phi_\nu=(constant,0,0,0)$ vector field may play the role of
cosmological constant \cite{dv1}. The more general case of
$\phi_\nu=(\alpha(t),0,0,0)$ has also interesting results in the
cosmological studies \cite{D1,dv2}. Recently, it has been
addressed that if one considers a suitable Lyra displacement
vector field, then some dynamics wormholes may satisfy
Eq.~(\ref{EinL}) in the FRW background \cite{dwhl}. In addition,
some exact solutions for the field equations in a Lyra manifold of
$\phi_\nu=(\alpha,0,0,0)$, where $\alpha$ is either a constant or
a function of $t$, have also been introduced which do not include
any wormhole \cite{s1,s2}. In fact, there is no study on
properties of a static traversable wormhole in the Lyra geometry,
and in most cases, physicist have focused on the cosmological
features of this theory
\cite{sen,D1,dv1,dv2,dwhl,s1,s2,hara,lyraobs,lkur,lrev,seng,lyrather,ln1,sepangi,ln2}.

As we have previously mentioned, the energy conditions are not met
by a traversable wormhole in GR \cite{cut-paste1}. In previous
section, we derived some new traversable wormholes with bounded
positive total mass. Thereinafter, we studied the effects of a
dark energy like source on the obtained solutions. In summary, our
investigation confirms that, in GR, the energy conditions are
violated by traversable wormholes, and a dark energy like source
cannot solve the problem of the energy condition violation.

Now, since WEC, NEC and DEC are independent of the gravitational
theory used to study the system, they lead to the same expression
for the energy density and pressure components as those obtained
in Eq.~(\ref{ec}). In order to find SEC in the Lyra manifold,
bearing the attractive nature of gravity in mind, we use the
Frobenius's theorem \cite{poisson} in order to find

\begin{eqnarray}\label{secl}
SECL\rightarrow\ \ \rho+p_r+2p_t\geq-3\phi_0\phi^0.
\end{eqnarray}

\noindent From now on, by SECL we mean the strong energy condition
in Lyra manifold.

Here, we also address some general remarks about the relation
between the anisotropy parameter and the Lyra displacement vector.
Using Eqs.~(\ref{an1}),~(\ref{EinL}) and~(\ref{lf}), one easily
finds

\begin{eqnarray}\label{anl}
\Delta&=&\mathcal{G}_2^2-\mathcal{G}_1^1=(G_2^2+\tilde{G}_2^2)-(G_1^1+\tilde{G}_1^1)\nonumber\\
&=&(G_2^2-G_1^1)+(\tilde{G}_2^2-\tilde{G}_1^1)=\Delta_E+\Delta_\phi,
\end{eqnarray}

\noindent where $\Delta_E\equiv G_2^2-G_1^1$ is the anisotropy
parameter corresponding to the Einstein tensor, and
$\Delta_\phi\equiv\frac{3}{2}[\phi_2\phi^2-\phi_1\phi^1]$ is due
to the Lyra displacement vector field. Therefore, for attractive
geometries $\Delta_E<-\Delta_\phi$, and in repulsive geometries
$\Delta_E>-\Delta_\phi$. It is also useful to note that whenever
$\Delta_E=0$, geometry is attractive (repulsive) if
$\phi_2\phi^2<\phi_1\phi^1$ ($\phi_2\phi^2>\phi_1\phi^1$).

\subsection{Asymptotically flat wormholes in Lyra manifold}

Considering $T_\mu^\nu=diag(-\rho,p_r,p_t,p_t)$ and by inserting
metric~(\ref{met}) into Eq.~(\ref{EinL}), we reach at

\begin{eqnarray}\label{denl0}
\rho&=&\frac{b^{\prime}(r)}{r^2}-\frac{3}{2}\phi_0\phi^0+\frac{3}{4}\phi_\alpha\phi^\alpha,\\
\nonumber p_r&=&\frac{U^{\prime}
(r)}{r U(r)}(1-\frac{b(r)}{r})-\frac{b(r)}{r^3}+\frac{3}{2}\phi_1\phi^1-\frac{3}{4}\phi_\alpha\phi^\alpha,\\
\nonumber
p_\theta&=&G_1^1+\frac{r}{2}\left[(G_1^1)^{\prime}+\left(G_1^1-G_0^0\right)\frac{U^{\prime}(r)}{2U(r)}\right]\\
\nonumber&+&\frac{3}{2}\phi_2\phi^2-\frac{3}{4}\phi_\alpha\phi^\alpha,\\
\nonumber
p_\phi&=&G_1^1+\frac{r}{2}\left[(G_1^1)^{\prime}+\left(G_1^1-G_0^0\right)\frac{U^{\prime}(r)}{2U(r)}\right]\\
\nonumber&+&\frac{3}{2}\phi_3\phi^3-\frac{3}{4}\phi_\alpha\phi^\alpha,
\end{eqnarray}

\noindent for the energy density and pressure components of the
supporter fluid. It is also useful to mention here that, due to
the spherically symmetric static nature of metric~(\ref{met})
implying, $p_\theta=p_\phi\equiv p_t$,
$\phi_2\phi^2=\phi_3\phi^3\equiv\gamma$, and $\phi_\mu$ should
either be constant or a function of radius. Therefore, the above
equations are written as

\begin{eqnarray}\label{denl}
\rho&=&\frac{b^{\prime}(r)}{r^2}-\frac{3}{2}\phi_0\phi^0+\frac{3}{4}\phi_\alpha\phi^\alpha,\\
\nonumber p_r&=&\frac{U^{\prime}
(r)}{r U(r)}(1-\frac{b(r)}{r})-\frac{b(r)}{r^3}+\frac{3}{2}\phi_1\phi^1-\frac{3}{4}\phi_\alpha\phi^\alpha,\\
\nonumber
p_t&=&G_1^1+\frac{r}{2}\left[(G_1^1)^{\prime}+\left(G_1^1-G_0^0\right)\frac{U^{\prime}(r)}{2U(r)}\right]\\
\nonumber&+&\frac{3}{2}\gamma-\frac{3}{4}\phi_\alpha\phi^\alpha.
\end{eqnarray}

As we saw in previous sections, radial pressure is negative at the
wormhole throat in GR. Now, based on these equations, one can
obtain that if
$(\frac{3}{2}\phi_1\phi^1-\frac{3}{4}\phi_\alpha\phi^\alpha)_{r=r_0}\geq\frac{1}{r_0^2}$,
then we have wormholes with $p(r_0)\geq0$. In this manner, if
$\rho(r_0)>0$, then, unlike GR, the $\rho+p_r>0$ condition is met
at the wormhole throat ($r=r_0$). Additionally, as we have
previously mentioned, since the Einstein and energy-momentum
tensors are diagonal here, at least three components of the Lyra
displacement vector field should be zero, a point will be
considered later.

Finally, using Eq.~(\ref{denl}), we easily reach at

\begin{eqnarray}\label{massfuncn}
m(r)&\equiv&\int_{r_0}^r
[\frac{b^{\prime}(r)}{r^2}-\frac{3}{2}\phi_0\phi^0+\frac{3}{4}\phi_\alpha\phi^\alpha]4\pi
r^2dr\\&=&4\pi(b(r)-b(r_0)-\frac{3}{2}\int_{r_0}^r
[\phi_0\phi^0-\frac{1}{2}\phi_\alpha\phi^\alpha] r^2dr),\nonumber
\end{eqnarray}

\noindent for the mass function.

In the following subsections, we find several traversable
wormholes and study some physical and mathematical properties of
the corresponding energy-momentum source in the Lyra manifold.

\subsection{The $U(r)=1$ case}\label{ff}

As the first example, let us consider the time-like Lyra
displacement vector field of $\phi_\mu=\beta\delta^t_\mu$ leading
to $\phi_\alpha\phi^\alpha=\phi_0\phi^0=-\beta^2$ and thus

\begin{eqnarray}\label{denl1}
\rho&=&\frac{b^{\prime}(r)}{r^2}+\frac{3}{4}\beta^2,\\
\nonumber p_r&=&-\frac{b(r)}{r^3}+\frac{3}{4}\beta^2,\\
\nonumber p_t&=&\frac{b(r)-rb^\prime(r)}{2r^3}+\frac{3}{4}\beta^2,
\end{eqnarray}

\noindent for the energy-momentum tensor. In obtaining these
results, one should remember that for $U(r)=1$, we have
$G^1_1=-\frac{b(r)}{r^3}$ (see also Eq.(\ref{den})). Inserting
these equations into Eq.~(\ref{secl}), one can easily obtain
$\rho+p_r+2p_t+3\phi_0\phi^0=0$ meaning that SECL is met. In order
to simplify our calculations, we define the transverse state
parameter as $w_t\equiv\frac{p_t}{\rho}$. Now, bearing the radial
and transverse state parameters in mind, and using the above
equations, we reach at

\begin{eqnarray}\label{denl2}
\beta^2&=&\frac{4}{3r^3(1-w_r)}[w_rrb^{\prime}(r)+b(r)],\\
\nonumber
\rho&=&\frac{1}{r^3(1-w_r)}[rb^{\prime}(r)+b(r)],\\
\nonumber p_r&=&\frac{w_r}{r^3(1-w_r)}[rb^{\prime}(r)+b(r)],\\
\nonumber
p_t&=&\frac{1}{2r^3(1-w_r)}[(3w_r-1)rb^{\prime}(r)+(3-w_r)b(r)],\\
\nonumber
w_t&=&\frac{1}{2}[\frac{3w_rrb^{\prime}(r)+(4-w_r)b(r)}{rb^{\prime}(r)+b(r)}-1].
\end{eqnarray}

The isotropic solutions are obtainable by applying the $w_r=w_t$
condition to the above equations leading to $b(r)=b_0r^3$, where
$b_0$ is the integration constant, which does not preserve the
$b(r)<r$ requirement for $r>r_0$. It is also apparent that this
solution does not respect both the $b(r_0)=r_0$ and the
flaring-out conditions simultaneously, meaning that this isotropic
solution is not a traversable wormhole.

\begin{figure}[ht] \centering
\includegraphics[width=2.5in, height=2.5in]{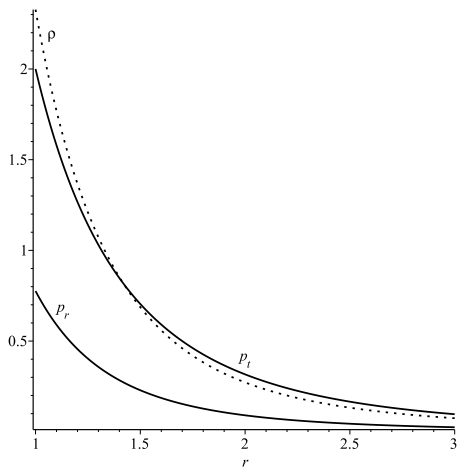}
\caption{\label{Lf1} The non-zero components of the
energy-momentum tensor for $b(r)=\frac{r_0}{\tanh(r_0)}\tanh(r)$,
$w_r=\frac{1}{3}$ and $r_0=1$. All curves tend to zero at the
$r\rightarrow\infty$ limit in accordance with the asymptotically
flat behavior of the metric. For this case, WEC, NEC and SECL are
satisfied, and DEC is not met everywhere.}
\end{figure}

Now, consider the $b(r)=\frac{r_0}{\tanh(r_0)}\tanh(r)$ case which
has extensively been studied in the GR framework in
Sec.~(\ref{11}). The evolution of the non-zero energy-momentum
components as functions of $r$ have been plotted for
$w_r=\frac{1}{3}$ and $r_0=1$ in Fig.~(\ref{Lf1}). As it is
obvious from both Eq.~(\ref{denl1}) and Fig.~(\ref{Lf1}), at the
$r\rightarrow\infty$ limit, the anisotropy of this repulsive
geometry ($\Delta>0$) is disappeared which is fully compatible
with the asymptotically flat behavior of the metric. It is also
apparent that this solution does not meet DEC everywhere. Here,
$p_t$ is positive and if we have a fluid with $0\leq w_r<1$, then
the radial pressure is positive. Moreover, since $r\geq r_0$, we
have
$\beta^2=\frac{2r_0}{3r^3\tanh(r_0)}[r(1-\tanh^2(r))+3\tanh(r)]>0$
and thus $\beta$ is a real quantity. Let us focus on the
$-1<w_r<1$ case for which the $\rho+p_r>0$ condition is
automatically satisfied in our example (because $\rho>0$). Now,
one can check that if the radial state parameter meets the
$-1<w_r<1$ condition, then a traversable wormhole with
$b(r)=\frac{r_0}{\tanh(r_0)}\tanh(r)$ and $U(r)=1$ satisfies NEC
and WEC expressed in Eq.~(\ref{ec}). In fact, only DEC is not
preserved by these solutions.

For the second example, considering the space-like Lyra
displacement vector field of $\phi_\mu=\o\delta^r_\mu$ which leads
to $\phi_\alpha\phi^\alpha=\phi_1\phi^1=\o^2(1-\frac{b(r)}{r})$,
and following the recipe which yielded Eq.~(\ref{denl2}), one
finds

\begin{eqnarray}\label{denl3}
\o^2&=&\frac{4}{3r^3(1-w_r)(1-\frac{b(r)}{r})}[w_rrb^{\prime}(r)+b(r)],\nonumber\\
\rho&=&\frac{1}{r^3(1-w_r)}[rb^{\prime}(r)+b(r)],\\
\nonumber p_r&=&\frac{w_r}{r^3(1-w_r)}[rb^{\prime}(r)+b(r)],\\
\nonumber
p_t&=&\frac{-1}{2r^3(1-w_r)}[rb^{\prime}(r)+w_rb(r)],\\
\nonumber
w_t&=&\frac{1}{2}[\frac{(1-w_r)b(r)}{rb^{\prime}(r)+b(r)}-1].
\end{eqnarray}

In this manner, since $\phi_0=0$, SECL is satisfied only if SEC is
met. Bearing the $b(r)=\frac{r_0}{\tanh(r_0)}\tanh(r)$ shape
function in mind, we can easily see that energy conditions
expressed in Eq.~(\ref{ec}) are met when $-1<w_r<1$. It is also
worthwhile mentioning that since for $r\geq r_0$ and $0\leq
w_r<1$, we have $\rho>0$, $p_r\geq0$, and $p_t<0$ (the transverse
pressure is negative), the geometry is attractive ($\Delta<0$).

In order to obtain the isotropic solutions corresponding to the
above space-like Lyra displacement vector, we apply the
$w_t=w_r=w$ condition to Eq.(\ref{denl3}) leading to

\begin{eqnarray}\label{Iso1}
\frac{db(r)}{b(r)}=-\frac{3wdr}{(2w+1)r},
\end{eqnarray}

\noindent which finally yields

\begin{eqnarray}\label{Iso2}
b(r)=r_0(\frac{r_0}{r})^{\frac{3w}{(2w+1)}},
\end{eqnarray}

\noindent for constant $w$. In obtaining the above equation, we
considered the $b(r_0)=r_0$ condition. Besides, using the
flaring-out condition, one can easily see that this condition is
met if $w$ either meets the $w\leq-\frac{1}{5}$ or the
$w>-\frac{1}{2}$ condition. For these values of $w$, we have
$b(r)<r$ for $r>r_0$, and thus $(1-b(r)/r)\rightarrow1$ at the
$r\rightarrow\infty$ limit meaning that metric preserves its
signature and the obtained solutions are asymptotically flat. It
is also useful to note here the $w=-\frac{1}{2}$ case leads to
$b(r)=0$ which is not a wormhole. Now, for the non-zero components
of energy-momentum tensor, we have

\begin{eqnarray}\label{denl4}
\rho&=&\frac{b(r)}{r^3(2w+1)},\\
\nonumber p&=&\frac{wb(r)}{r^3(2w+1)},
\end{eqnarray}

\noindent where $p=p_t=p_r$ is the isotropic pressure. In order to
evaluate the mass function of the obtained solution, one can use
Eq.~(\ref{massfuncn}) to get

\begin{eqnarray}
m(r)=\frac{4\pi r_0}{3w}[1-(\frac{r_0}{r})^\frac{3w}{(2w+1)}],
\end{eqnarray}

\noindent for $w\neq0$ and therefore, mass is positive and bounded
for these solutions if we have $w>0$. For the $w=0$ case, simple
calculation leads to $m(r)=4\pi r_0\ln(\frac{r}{r_0})$ meaning
that mass increases as a function of $r$, and therefore, it is a
solution with unbounded mass. In fact, for $w<-\frac{1}{2}$ and
$0<w$, mass is bounded (or equally $m\rightarrow\frac{4\pi
r_0}{3w}$ at the $r\rightarrow\infty$ limit), and it is unbounded
for $-\frac{1}{2}<w<0$. Combining these results with
Eq.~(\ref{denl4}), one can easily see that fluids with state
parameter meeting the $-\frac{1}{3}\leq w$ condition satisfy
energy conditions expressed in Eq.~(\ref{ec}), and therefore, the
energy conditions are satisfied by the baryonic energy-momentum
source (or equally $0\leq w\leq\frac{1}{3}$).

In summary, we obtained that if a suitable space-like Lyra
displacement vector is chosen, then the isotropic baryonic sources
can theoretically support the traversable asymptotically flat
wormholes in the Lyra manifold.

\subsection{The exponential redshift function}

Here, we focus on the $U(r)=\exp(cr^n)$ case, which led to
interesting results in Sec.~(\ref{nn}). Moreover, since this case
is a generalization of Sec.~(\ref{ff}), we only consider the
time-like Lyra displacement vector fields. As a reminder, we found
that, in GR framework~(\ref{nn}), $n<0$ leads to asymptotically
flat wormholes that for them radial pressure may get positive
values only if $c<0$ (see Eq.~(\ref{c1})). Therefore, we only
consider the $n,\ c<0$ case. Considering the time-like vector of
$\phi_\mu=\Sigma\delta^t_\mu$ and using~(\ref{denl}), one obtains

\begin{eqnarray}\label{denl6}
\rho&=&\frac{p_r}{w_r}=\frac{1}{r^3(1-w_r)}[rb^{\prime}(r)+b(r)\nonumber\\&-&cnr^{n+1}(1-\frac{b(r)}{r})],\\
\nonumber
w_t&=&\frac{w_r-1}{4(rb^{\prime}(r)+b(r)-cnr^{n+1}(1-\frac{b(r)}{r}))}[rb^{\prime}(r)\nonumber\\&\times&(2+cnr^n)-cn^2r^{n+1}(1-\frac{b(r)}{r})(2+cr^n)\nonumber\\
&+&2b(r)(cnr^{n-1}-1)+\frac{4}{w_r-1}[w_rrb^{\prime}(r)\nonumber
\\&+&b(r)-cnr^{n+1}(1-\frac{b(r)}{r})]],\nonumber
\end{eqnarray}

\noindent whenever
$\Sigma^2=\frac{4}{3(1-w_r)\exp(cr^n)}[\frac{w_rb^{\prime}(r)}{r^2}+\frac{b(r)}{r^3}(1+cnr^n)-cnr^{n-2}]$.
Inserting $w_t=w_r=w$ in the above equation to find the isotropic
solutions, one reaches at $w=1$ which is the state parameter of
stiff matter. Now, let us focus on the
$b(r)=\frac{r_0\tanh(r)}{\tanh(r_0)}$ case. In this manner,
although we have $\rho>0$ and $p_r\geq0$ for $0\leq w_r<1$, the
$\rho+p_t\geq0$ condition is not satisfied for all values of $r$
bigger than of $r_0$.

For another example, considering the $\phi_\mu=\Theta\delta^t_\mu$
case, where $\Theta^2=\frac{4b(r)}{3U(r)r^3}$, we can use
Eq.~(\ref{denl}) in order to get

\begin{eqnarray}\label{enf1}
\rho&=&\frac{r_0[1-\tanh^2(r)]}{r^2\tanh(r_0)}+\frac{r_0\tanh(r)}{\tanh(r_0)r^3},\\
p_r&=&cnr^{n-2}(1-\frac{r_0\tanh(r)}{\tanh(r_0)r}),\nonumber\\
\nonumber
p_t&=&\frac{1}{4r^3\tanh(r_0)}[(cnr^n)^2(r\tanh(r_0)-r_0\tanh(r))\nonumber\\
&+&cnr^n(r_0(r\tanh^2(r)+(1-2n)\tanh(r))\nonumber\\
&+&r(2n\tanh(r_0)-r_0))+\tanh(r))\nonumber\\
&+&2r_0(r(\tanh^2(r)-1)]+\frac{r_0\tanh(r)}{\tanh(r_0)r^3}.\nonumber
\end{eqnarray}

As it is obvious, the radial pressure and energy density are
positive quantities for $r\geq r_0$. Moreover, comparing these
results with those obtained in Eq.~(\ref{en1}), one can easily
obtain that $p_t$ is also positive for $r\geq r_0$. Therefore,
while WEC and NEC are satisfied by this solution, DEC is not met
everywhere. In addition, it is easy to check that SECL is also
satisfied by this case. These results are similar to those
obtained in Sec.~(\ref{ff}) for time-like case.

\section{Traversable wormholes in an empty Lyra manifold}

In this section, we are going to study the wormhole solutions of
an empty Lyra manifold. In this manner, $T_{\mu\nu}=0$ and we have

\begin{eqnarray}\label{EinLf}
G_{\mu\nu}&=&\frac{3}{4}\phi_\alpha\phi^\alpha
g_{\mu\nu}-\frac{3}{2}\phi_\mu\phi_\nu,\nonumber\\
R&=&-\frac{3}{2}\phi_\alpha\phi^\alpha.
\end{eqnarray}

\noindent where $R$ is the Ricci scalar. From this equation, it is
obvious that the $\phi_\mu=0$ case, yielding $G_{\mu\nu}=0$, does
not lead to a traversable wormhole. One can also abridge the above
equations as

\begin{eqnarray}\label{EinLf0}
R_{\mu\nu}&=&-\frac{3}{2}\phi_\mu\phi_\nu.
\end{eqnarray}

\noindent Here, $R_{\mu\nu}$ denotes the Ricci tensor. Therefore,
for spacetimes in which the Ricci tensor is diagonal (or equally
$R_\mu^\nu\propto\delta_\mu^\nu$), only one component of the
$\phi_\mu\phi_\nu$ tensor can be non-zero. It means that only one
component of $R_{\mu\nu}$ can be non-zero unless we have
$\phi_\mu=0$ leading to $R_{\mu\nu}=0$ and thus $G_{\mu\nu}=0$.

For an observer with four velocity $v_\mu$, WEC is defined as
$T_{\mu\nu} v^\mu v^\nu\geq0$ \cite{poisson}. Now, using
Eq.~(\ref{EinL}), one reaches at $\mathcal{G}_{\mu\nu}v^\mu
v^\nu\geq0$ for WEC. In fact, it is the geometrical interpretation
for WEC in the Lyra manifold. For an empty Lyra manifold, where
$\mathcal{G}_{\mu\nu}=T_{\mu\nu}=0$, we have
$\mathcal{G}_{\mu\nu}v^\mu v^\nu=0$ meaning that WEC is marginally
satisfied in an empty Lyra manifold. One can use the definitions
of NEC and DEC \cite{poisson} to check that similar argument is
also valid for them in an empty Lyra manifold. Therefore,
independent of the Lyra displacement vector field, WEC, NEC and
DEC are marginally satisfied in an empty Lyra manifold. Finally,
bearing Eq.~(\ref{secl}) in mind, we easily find that SECL is met
if $3\phi_0\phi^0\geq0$.


\subsection{Time-like displacement vector field\label{tim}}

In our signature ($-,+,+,+$), $\phi_\mu$ is a time-like vector
field if $\phi_\mu\phi^\mu<0$. Therefore, a time-like vector as
$\phi_\mu=\zeta(r)\delta^t_\mu$ should respect the
$\phi_\mu\phi^\mu=-\frac{\zeta^2(r)}{U(r)}<0$ condition. Now,
considering $\phi_\mu=\zeta(r)\delta^t_\mu$ and using
metric~(\ref{met}) as well as Eq.~(\ref{EinLf0}), we obtain

\begin{eqnarray}\label{ef}
\zeta(r)=\sqrt{\frac{2R^0_0U(r)}{3}},
\end{eqnarray}

\noindent where

\begin{eqnarray}\label{zero}
&R^0_0=\frac{1}{4r^2U(r)^2}[-2rU(r)U^{\prime\prime}(r)(r-b(r))\\
&+(rU^{\prime}(r))^2(1-\frac{b(r)}{r})+U(r)U^{\prime}(r)(3b(r)+rb^\prime(r)-4r)]\nonumber
\end{eqnarray}

\noindent In addition, from Eq.~(\ref{EinLf}), one also finds

\begin{eqnarray}\label{ef0}
\zeta(r)=\sqrt{\frac{2RU(r)}{3}},
\end{eqnarray}

\noindent equal with Eq.~(\ref{ef}), only if we have $R=R^0_0$ and
thus $R^1_1=R^2_2=R^3_3=0$. The $R^2_2=R^3_3=0$ condition leads to

\begin{eqnarray}\label{ef1}
\frac{U^\prime(r)}{U(r)}=\frac{b(r)+rb^\prime(r)}{r^2-rb(r)},
\end{eqnarray}

\noindent which can finally be written as

\begin{eqnarray}\label{ef2}
\frac{U^{\prime\prime}(r)}{U(r)}=\frac{r^2(1-\frac{b(r)}{r})b^{\prime\prime}(r)-2(1+b^\prime(r))b(r)}{r(r-b(r))^2}.
\end{eqnarray}

\noindent Now, inserting Eqs.~(\ref{ef1}) and~(\ref{ef2}) into the
$R^1_1=0$ condition, one can reach an equation for $b(r)$ as

\begin{eqnarray}\label{ef3}
b^{\prime\prime}(r)=\frac{r^2b^\prime(r)(2+b^\prime(r))+b(r)(2b(r)+rb^\prime(r))}{r^3(1-\frac{b(r)}{r})}.
\end{eqnarray}

Therefore, if the shape function of a traversable wormhole
satisfies this equation, then its redshift function may be
evaluated from Eq.~(\ref{ef1}). In this situation, the time-like
Lyra displacement vector, supporting the traversable wormhole in
the empty Lyra geometry, meets Eq.~(\ref{ef}). Inserting
Eq.~(\ref{ef3}) as well as Eqs.~(\ref{ef2}) and~(\ref{ef1}) into
Eq.~(\ref{zero}), we finally reach at

\begin{eqnarray}\label{zero1}
&R^0_0=\frac{-2b^\prime(r)(1-\frac{5b(r)}{4r})}{r^2(1-\frac{b(r)}{r})}.
\end{eqnarray}

Indeed, based on the above equations, if one of the $b(r)$, $U(r)$
and $\zeta(r)$ functions is known, then the two remaining
functions may be evaluated. The only important thing is that
Eqs.~(\ref{ef0}-\ref{ef3}) should be respected by the solutions.
It is also useful to remind here that, in order to have a
traversable wormhole, the obtained solutions for $b(r)$ and $U(r)$
should meet the requirements of a traversable wormhole expressed
in Sec.~(\ref{ter}).

As we have previously claimed, SECL is met if $3\phi_0\phi^0\geq0$
which yields the condition $\zeta^2(r)\leq0$. On the other hand,
we have $\zeta^2(r)>0$ for a time-like Lyra displacement vector
field. Therefore, a time-like displacement vector cannot support
the obtained geometries and SECL in an empty Lyra manifold
simultaneously. In summary, if we consider the mentioned time-like
vector field and use the above recipe to find some traversable
wormholes, solutions will not respect SECL.

Considering $U(r)=1$, and by using Eq.~(\ref{ef1}), we get
$b(r)=\frac{r_0^2}{r}$, and thus
$\zeta(r)=\sqrt{\frac{8r_0^2(r^2-\frac{5r_0^2}{4})}{3r^3(r^2-r_0^2)}}$.
In order to obtain the shape function, the $b(r_0)=r_0$ has also
been used, and it is obvious that it meets the flaring-out
condition. Moreover, for $r\geq\sqrt{\frac{5}{4}}r_0$ we have
$\zeta^2(r)>0$ and thus SECL is not met. We should also note that
SECL is met ($\zeta^2(r)<0$) for $r_0\leq
r<\sqrt{\frac{5}{4}}r_0$, but in this manner, the Lyra
displacement vector field is imaginary. Although the physical
meaning of an imaginary field is not clear \cite{lrev}, some
physical consequences of this field and its mathematical
properties have been studied, found in Ref.~\cite{lrev} and
references therein.

Now, let us consider the $U(r)=\exp(cr^n)$ case. Using
Eq.~(\ref{ef1}), we get $b(r)=\frac{r_0^2}{r}\exp(-c[r^n-r_0^n])$
and thus
$\zeta(r)=\sqrt{\frac{8r_0^2(1+cnr^{n-1})\exp(cr_0^n)(r^2-\frac{5r_0^2}{4}\exp(-c[r^n-r_0^n]))}{3r^3(r^2-r_0^2\exp(-c[r^n-r_0^n]))}}$,
where the $b(r_0)=r_0$ condition have also been applied to the
solution. The flaring-out condition is also met if we have
$c\geq-\frac{r_0+1}{nr_0^n}$ leading to this fact that the $n,c<0$
case does not respect the flaring-out condition. Moreover, the
$b(r)<r$ condition, for $r>r_0$, is available if we have $n,c>0$
or $c>0$ for $n<0$ meaning that $b(r)$ is a decreasing function of
$r$ with a maximum located at $r=r_0$. In addition, whenever
$r^2\exp(cr^n)\geq\frac{5r_0^2}{4}\exp(cr_0^n)$, we have
$\zeta(r)\geq0$ and as a result, SECL is not preserved.

In summary, we found out that, in an empty Lyra manifold, a real
time-like vector field, parallel to the $R^0_0>0$ case, cannot
lead to a traversable wormhole which satisfies SECL.

\subsection{Space-like displacement vector fields}\label{ff1}

For the space-like displacement vector fields, whenever
$G_{\mu\nu}$ and $T_{\mu\nu}$ are diagonal, only one spatial
component of the Lyra displacement vector field is non-zero, and
moreover, since $\phi_0=0$, SECL is automatically preserved.

\subsection*{Transverse space-like vector fields}

For the $\phi_\mu=\phi\delta_\mu^\theta$ case, on one hand,
following the recipe of the previous subsection, we get the
$R_3^3=0$ and $\phi=\sqrt{\frac{-2r^2R^2_2}{3}}$ conditions. On
the other hand, due to the spherical symmetry of the assumed
static metric~(\ref{met}), $R_2^2=R_3^3$ and $\phi$ should either
be a constant or a function of $r$. Therefore, we should have
$\phi(r)=0$ meaning that such vector field does not exist. Similar
argument is also available for the
$\omega_\mu=\omega\delta_\mu^\phi$ case and thus, a Lyra manifold
with $\omega_\mu=\omega\delta_\mu^\phi$, where $\omega$ is a
constant or a function of $r$, cannot satisfy the field equations.

\subsection*{Radial Space-like vector field}
Now, let us consider the $\phi_\mu=\xi(r)\delta_\mu^r$ case. From
Eq.~(\ref{EinLf0}), we have

\begin{eqnarray}\label{ef4}
\xi(r)=\sqrt{\frac{-2R^1_1}{3(1-\frac{b(r)}{r})}}=\sqrt{\frac{-2R}{3(1-\frac{b(r)}{r})}},
\end{eqnarray}

\noindent which is true only if $R^0_0=R^2_2=R^3_3=0$. The
$R^2_2=R^3_3=0$ condition leads again to Eqs.~(\ref{ef1})
and~(\ref{ef2}) combined with $R^0_0=0$ to reach at

\begin{eqnarray}\label{ef5}
b^{\prime\prime}(r)&=&\frac{1}{r^3(r-b(r))}[b^\prime(r)(r^2(b^\prime(r)-2)\nonumber\\&+&b(r)(3r-1))+b(r)(b(r)-2r)].
\end{eqnarray}

\noindent Finally, it is a matter of calculation to show

\begin{eqnarray}\label{ef6}
\frac{U^{\prime\prime}(r)}{U(r)}&=&\frac{1}{r^3(r-b(r))^2}[b^\prime(r)(r^2(b^\prime(r)-2)\\&+&b(r)(r(3-2r)-1))+b(r)(b(r)-2r(1+r))].\nonumber
\end{eqnarray}

In fact, the above results help us in finding proper $\xi(r)$.
Therefore, once the shape function is known, we can use the above
equations to find the redshift function and the Lyra displacement
vector field. Theoretically, if one of the $\xi(r)$, $b(r)$ or
$U(r)$ functions is known, then one can use the above equations in
order to find the two remaining functions.

As an example, we consider the simple case of $U(r)=1$. We only
focus on this simple case because the exponential redshift
function leads to boring functions for the shape function which
make analysis hard and impossible in some steps. In fact, this
simple case is enough to show that SECL can be satisfied by
traversable wormholes in an empty Lyra manifold, if a proper
radial space-like Lyra displacement vector has been chosen. In
this manner, inserting metric~(\ref{met}) into Eq.~(\ref{EinLf0}),
one reaches at

\begin{eqnarray}\label{Einlf1}
&&\frac{3}{2}\phi_0\phi^0=0,\\
&&R_1^1=-\frac{3}{2}\phi_1\phi^1=\frac{rb^\prime(r)-b(r)}{r^3},\nonumber\\
&&R_2^2=R_3^3=-\frac{3}{2}\phi_2\phi^2=-\frac{3}{2}\phi_3\phi^3=\frac{b(r)+rb^\prime(r)}{2r^3}.\nonumber
\end{eqnarray}

\noindent It is obvious that a time-like vector as
$\phi_\mu=\xi\delta^t_\mu$, where $\xi\neq0$, cannot satisfy the
above equations, a result which is in full agreement with our
results obtained in subsection~(\ref{tim}).

The above equations are valid only for $b(r)=\frac{b_0}{r}$, where
$b_0$ is the integration constant, a shape function obtained and
studied in a dynamics background \cite{dwhl}. The $b(r_0)=r_0$
condition lead to $b(r)=\frac{r_0^2}{r}$ and thus

\begin{eqnarray}\label{Einlf2}
&&\phi_1\phi^1=-\frac{2}{3}R=\frac{4r_0^2}{3r^4},\\
&&\phi_0=\phi_2=\phi_3=0.
\end{eqnarray}

It is also easy to check that the flaring-out condition is
available here ($b^\prime(r_0)=-1<1$). Therefore, the vacuum Lyra
manifold with a space-like displacement vector filed as
$\phi_\mu=\frac{2r_0}{r^2\sqrt{3(1-\frac{r_0^2}{r^2})}}\delta^r_\mu$
may support a traversable asymptotically flat wormhole. Finally,
we should remind that since we have $\phi_0=0$, SECL is
automatically satisfied here.

\section{Conclusion}

Firstly, we introduced some new traversable wormholes and studied
their physical properties in the GR framework. In addition, the
asymptotically flat case as well as the effects of considering a
source with energy-momentum similar to the cosmological constant
on the obtained solutions have also been addressed.

Thereinafter, providing a summary on the Einstein field equations
in the ordinary gauge of the Lyra manifold as a generalization of
the Reimannian geometry, we studied some properties of the
wormhole's structure in the Lyra geometry. In Sec.~(\ref{ff}), we
derived some isotropic traversable wormholes which are
asymptotically flat and can meet energy conditions if a proper
space-like Lyra displacement vector field is chosen. Therefore,
our study shows that baryonic matters are allowed to support
traversable wormholes and also the asymptotically flat cases in
the Lyra manifold.

Finally, considering an empty spacetime, the possibility of having
traversable wormholes in the empty Lyra manifold has been
investigated. Results show that the empty Lyra manifold may
support traversable wormholes and also the asymptotically flat
cases. Additionally, focusing on the simple case of $U(r)=1$ in
Sec.~(\ref{ff1}), we derived a traversable asymptotically flat
wormhole in an empty Lyra manifold for that energy conditions are
marginally satisfied if a proper radial space-like Lyra
displacement vector is chosen. Therefore, it is theoretically
possible to choose a suitable Lyra displacement vector field which
lets the empty Lyra manifold support traversable wormholes.

\section*{Acknowledgments}
We are so grateful to the anonymous referee for valuable comments.
The work of H. Moradpour has been supported financially by
Research Institute for Astronomy \& Astrophysics of Maragha
(RIAAM) under project No.1/4717-171.

\end{document}